\documentstyle[multicol,aps,prb,epsf]{revtex}
\begin{document}

\draft

\title{
Electronic properties of 
metal induced gap states at insulator/metal interfaces\\
--- dependence on the alkali halide and the possibility of 
excitonic mechanism of superconductivity
}

\author{
Ryotaro Arita$^1$, Yoshiaki Tanida$^2$, Kazuhiko Kuroki$^3$, 
and Hideo Aoki$^1$
}

\address{$^1$Department of Physics, University of Tokyo, Hongo,
Tokyo 113-0033, Japan}
\address{$^2$Fujitsu Laboratories Ltd., Atsugi, Kanagawa 243-0197, Japan}
\address{$^3$Department of Applied Physics and Chemistry,
University of Electro-Communications, Chofu, Tokyo 182-8585, Japan}

\date{\today}

\maketitle

\begin{abstract}
Motivated from the experimental observation of metal induced 
gap states (MIGS) at insulator/metal interfaces by 
Kiguchi {\it et al.} [Phys. Rev. Lett. {\bf 90}, 196803 (2003)], 
we have theoretically investigated the electronic properties of 
MIGS at interfaces between various alkali halides and a metal 
represented by a jellium with the 
first-principles density functional method.
We have found that, on top of the usual evanescent state, 
MIGS generally have a long tail on halogen sites with a 
$p_z$-like character, whose penetration depth ($\lambda$) 
is as large as half the lattice constant of bulk alkali halides.  
This implies that $\lambda$, while little dependent on the carrier 
density in the jellium, is dominated by the lattice constant 
(hence by energy gap) of the alkali halide, 
where $\lambda_{\rm LiF} < \lambda_{\rm LiCl} < \lambda_{\rm LiI}$.
We also propose a possibility of the MIGS working favorably 
for the exciton-mediated superconductivity.
\end{abstract}

\medskip


\begin{multicols}{2}
\narrowtext

\section{Introduction}
Heterointerfaces, namely solid-solid interfaces between 
very dissimilar materials, have been an issue of great 
interest in condensed matter physics, where 
various fascinating possibilities may be considered.
Especially, the electronic structure of metal/insulator 
or metal/semiconductor interfaces has a long 
history of intensive studies dating back to 
the 1930's\cite{msreview,mireview}. 

One of the most crucial factors which govern the
electronic structure of heterointerfaces is 
the metal-induced gap states (MIGS), 
whose eigenenergies lie in the band gap of the semiconductor 
(or insulator).  While MIGS spread well into the metal side 
of the interface, they decay exponentially into the semiconductor.  
The decay length is an order of a few \AA\ for typical 
semiconductors, where MIGS have been believed to determine the 
Schottky barrier 
height\cite{cohen75,cohen76,cohen77,Tersoff84,Tersoff86,Berthod03}. 

If we have an insulator/metal interface rather than a 
semiconductor/metal interface, 
on the other hand, it has generally been believed that 
the decay length of the MIGS into the insulator is negligible.  
This view has been based on 
quite a plausible model calculation, where the
insulator/metal is represented by a simple tight-binding 
model/jellium\cite{Noguera}.  
For a typical ionic insulator such as LiCl, the band gap ($\sim O(10)$ eV) 
is an order of magnitude greater than the transfer energy ($\sim O(1)$ eV) 
between the cations and anions, and 
the decay length is estimated to be smaller than 1 \AA,
which would suggest that MIGS should be 
irrelevant at insulator/metal interfaces. 

Experimentally, 
it is difficult to detect MIGS at the heterointerface
between wide gap insulators and metal, because clean interfaces 
are difficult to be fabricated. 
While MIGS have been reported in MgO/Cu(222)\cite{muller98}
or MgO/Ag(001)\cite{Schintke01}, the question of 
whether the observed gap 
states are formed by local chemical bonding remained,
since MgO(222) is a polar surface so that a 
strong hybridization between the O 2p band and the Ag 5sp 
band is expected for MgO/Ag\cite{Altieri99}.

Recently, Kiguchi {\it et al}\cite{Kiguchi} have succeeded in 
observing MIGS at LiCl/Cu and LiCl/Ag interfaces. They have 
grown a LiCl film on a metal substrate in a layer-by-layer 
fashion, and exploiting 
the near edge x-ray absorption fine structure (NEXAFS) 
which is an element-selective method, they have
observed an extra peak below the bulk edge onset for LiCl film,
which was interpreted as an evidence for MIGS. 
The reason why the detection eluded earlier studies is 
that, in conventional methods such as ultraviolet photoemission 
spectroscopy(UPS), inverse photoemission spectroscopy, 
or electron energy loss spectroscopy(EELS), 
the signal from the interface is obscured by
significant contributions from the substrate. 
Kiguchi et al. have estimated 
the decay length of MIGS, which are 
$2.6\pm 0.3$ \AA\ for LiCl/Cu(001) and $2.9\pm 0.7$ \AA\ for 
LiCl/Ag(001), values much 
greater than those expected in simple models\cite{Noguera}. 
An {\it ab initio} electronic structure calculation 
for LiCl/metal system has 
supported the existence of the MIGS that is strongly 
localized at the interface\cite{Kiguchi}.

The purpose of the present paper is to study the
electronic structure of interfaces between various alkali halides and a
metal by means of first principles calculation, and to 
search for novel possibilities specific to metal/insulator interfaces.
We shall show that (i) MIGS have a tail on halogen sites with a 
$p_z$-like character, whose penetration depth ($\lambda$) 
is as large as half the lattice constant of the bulk alkali halide,
which are totally consistent with the experimental 
result\cite{Kiguchi}, (ii) while $\lambda$ is insensitive to the 
carrier density in the metal, the penetration depth is 
dominated by the energy gap of the alkali halide.  
Specifically, the band gaps are 
$E_{\rm LiF} > E_{\rm LiCl} > E_{\rm LiI}$, and the 
the penetration depth has the expected inequality, 
$\lambda_{\rm LiF} < \lambda_{\rm LiCl} < \lambda_{\rm LiI}$.

We can consider various possibilities for many-body 
effects unique to heterointerfaces.  
There is in fact a long history for those, 
where the notable ones include 
metal-insulator transition\cite{Anderson} and 
superconductivity\cite{Little64,Ginzburg64,Ginzburgbook,ABB}. 
In the context of the latter, here we propose that 
large density of states arising from the MIGS inside the insulator 
is a good news for the superconductivity due to the exciton mechanism.   
The history of the exciton mechanism of superconductivity 
dates back to the proposals by Little\cite{Little64},
Ginzburg\cite{Ginzburg64,Ginzburgbook}, and Bardeen {\it et al}\cite{ABB}.
For this mechanism, the coexistence, in real space, of dispersionless
excitons and metallic carriers is required\cite{ABB}. 
While this is difficult 
to achieve in general, if the density of states due to MIGS inside the
insulator is large enough,
we may envisage that the insulator/metal interface such as those 
discussed here may provide a possible ground for 
superconductivity. We shall discuss this toward the end of the paper.

\section{method}
The electronic structure of MIGS is studied
with a first-principles calculation based on the
density functional theory.
We adopt the exchange-correlation functional introduced by
Perdew, Burke and Wang\cite{Perdew1996} and ultra-soft 
pseudo-potentials\cite{Vanderbilt90,Laasonen93} in a separable form.
We have first calculated the bulk system and determined the
lattices. The obtained values 
(against experimental ones in parentheses) are 4.09 (4.02) \AA\ 
for LiF, 5.13 (5.13) \AA\ for LiCl, 5.95 (6.00) \AA\ 
for LiI and 5.63 (5.63) \AA\ for NaCl.

\begin{figure}
\begin{center}
\leavevmode\epsfysize=60mm \epsfbox{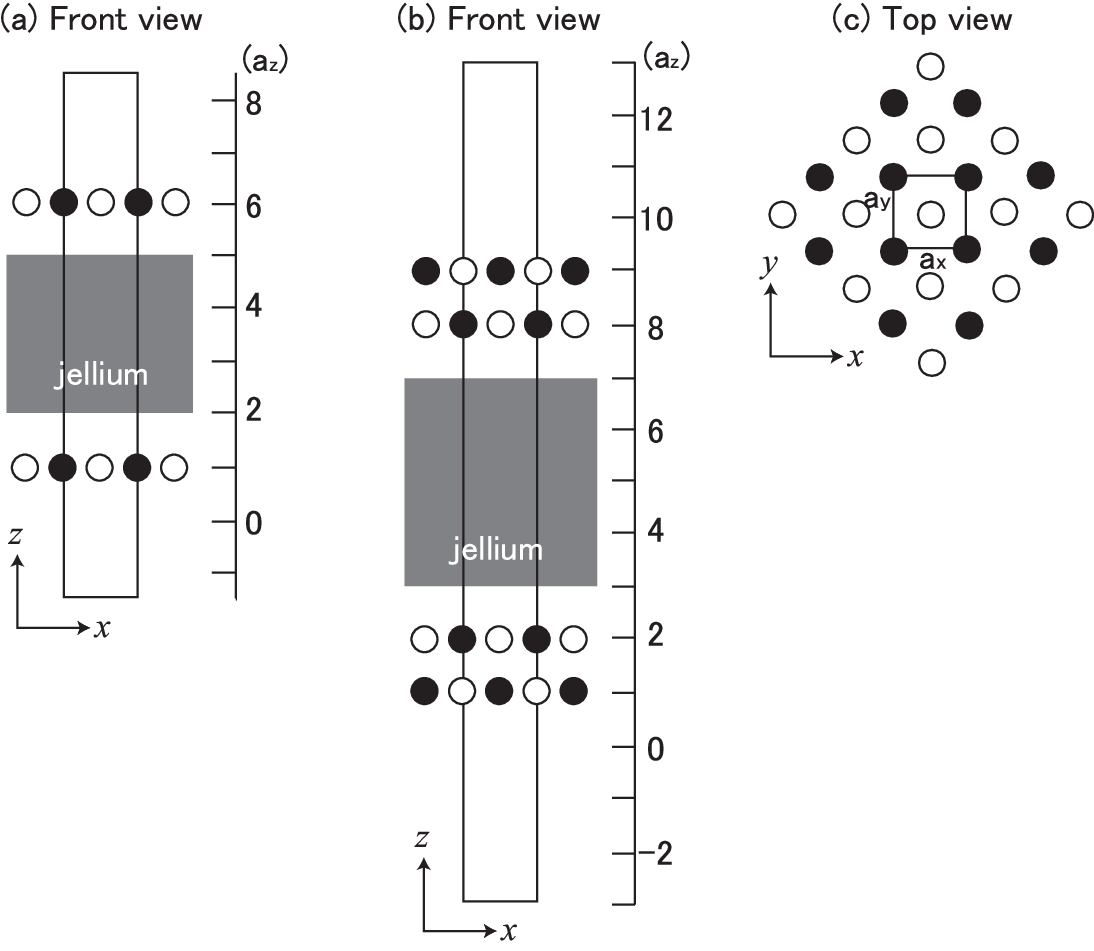}
\caption{Front(a,b) and top(c) views 
of the atomic configuration considered in the present study.
In a(b) one (two) layer(s) of alkali halide 
sandwich a metal region modeled by a jellium, where 
solid (open) circles denote cation (anion) atoms while 
shaded regions the jellium. Unit cells are indicated. 
The lattice constant along $x$, $y$ directions 
is denoted as $a_x=a_y=a/\sqrt{2}$, 
where $a$ is the lattice constant defined 
for the bulk alkali halide with NaCl structure.  
The interlayer distance is denoted as $a_z=a/2$. 
}
\label{unit}
\end{center}
\end{figure}

We have then introduced a slab model, where
we put a jellium in the middle of the unit cell 
sandwiched from top and bottom by an alkali halide layer
(Fig.\ref{unit}).  
Starting from this structure as an initial state, 
we have performed structure optimization.
To confirm that the simplified model with 
the jellium capture main features in heterostructures 
with real metals, we have also studied the system where
five layers of Cu is considered in place of the jellium.
The lattice mismatch between LiCl and Cu happens to be 
small, so that we can take a small unit cell.
The cut-off energy of the plane-wave expansion for
the wave functions are 20.25 Ry for LiF, LiCl and LiI/jellium, 
25.00 Ry for NaCl/jellium, and 56.25 Ry for LiCl/Cu.
The atomic configurations and the corresponding electronic states
in the ground states are obtained with the conjugate gradient
scheme\cite{Yamauchi1996}.

The local density of states is 
calculated as $\sum_{i}|\phi_i(x,y,z)|^2$,
where $\phi_i$'s are eigenfunctions and the summation is taken over 
a window $E_F<E_i<E_F+0.5 {\rm eV}$ with $E_F$ being the Fermi energy.  
We have set the window above $E_F$ 
to compare with the experiment, where the NEXAFS\cite{Kiguchi}
detects the unoccupied states. 

\section{Results}
\subsection{LiCl}
Let us first discuss the case of LiCl/jellium.
Since MIGS have been observed both in LiCl/Cu(commensurate) 
and LiCl/Ag(incommensurate)\cite{Kiguchi},
the commensurability in the lattice constants across the 
heterointerface should not be 
essential for the formation of MIGS, 
as we shall confirm below from the results for 
LiCl/Cu and LiCl/jellium. So we start with the jellium model.

\begin{figure}
\begin{center}
\leavevmode\epsfysize=70mm \epsfbox{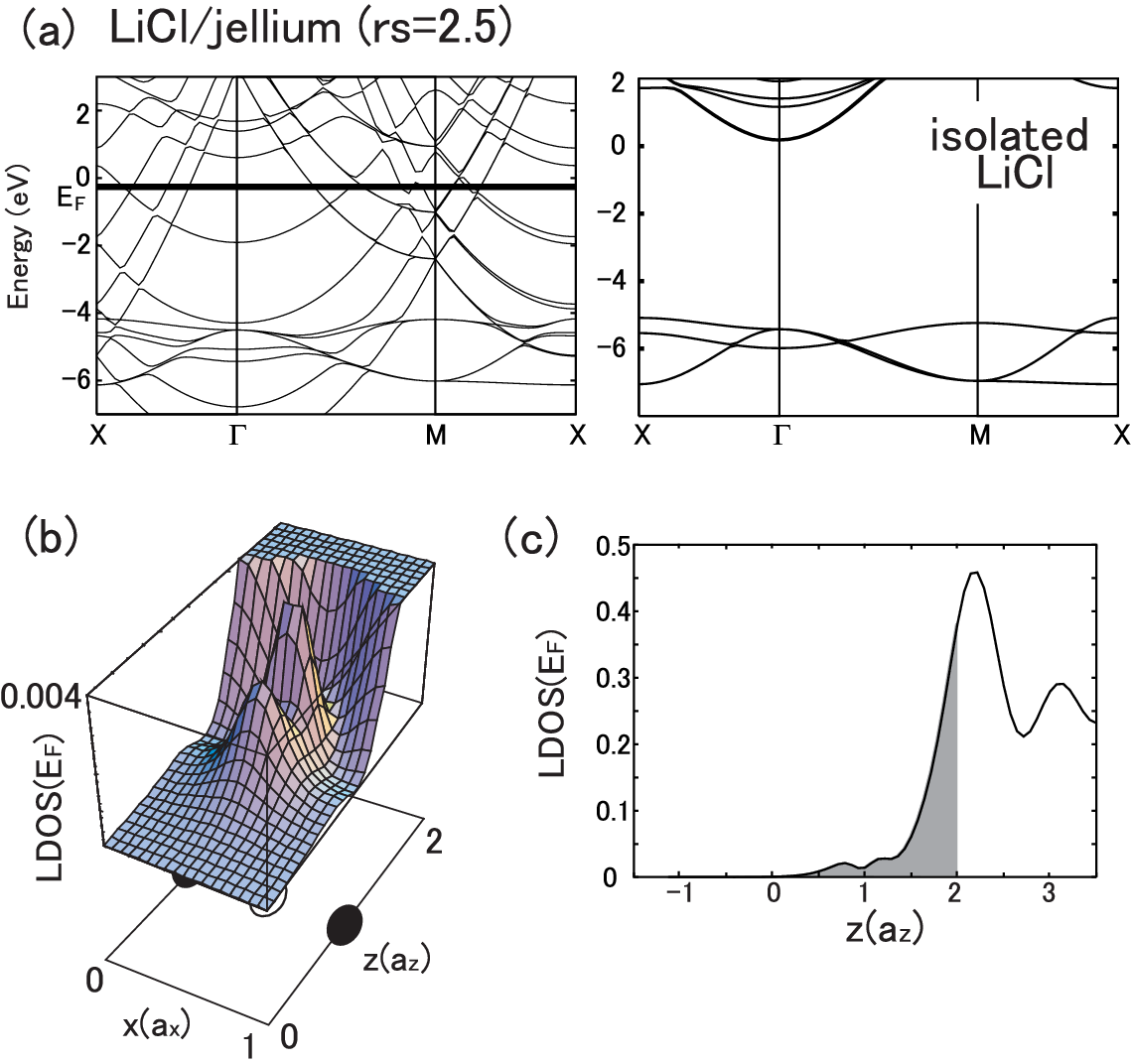}
\caption{(a) Band structure for the system where
LiCl monolayers sandwich the jellium having $r_s=2.5$ (left panel)
and that for an isolated LiCl (i.e., $r_s=\infty$, right panel).
(b) Local density of states (LDOS) for $E_F <E <E_F+0.5 {\rm eV}$ 
integrated over $y$-direction. The solid (open) circle
denotes Li (Cl).
(c) LDOS integrated over both $x$ and $y$ directions. 
The shaded region indicates the insulator region.
}
\label{rs25}
\end{center}
\end{figure}

We show in Fig.\ref{rs25}(a) the band structure of 
LiCl/jellium where LiCl layers, each one monolayer thick, 
sandwiches the jellium that has $r_s=2.5$. 
We have also displayed the band structure for an isolated LiCl monolayer. 
From the comparison we can see that the band structure of LiCl is not severely
affected by the presence of the jellium.

It is well-known that GGA or 
LDA generally underestimate the band gap, while 
GW approximation\cite{gunnarsson98,Aulbur00} improves this.  
For instance, the band gap for bulk LiCl is 9.2 eV 
in GW, which is very close to 
the experimental value of 9.4 eV\cite{Shirley98}.
However, the band-gap underestimation may be amended 
via the self-energy correction (i.e., without any
corrections for LDA wave functions), 
the shape of the wave functions should be reliable even in 
LDA, hence the present calculation
for LDOS around the $E_F$ is expected to be a good approximation.

In Fig.\ref{rs25}(b,c) we have plotted the local density of states (LDOS)
integrated along $x$ or $y$-directions.
We can see that LDOS has a significant 
($\sim$ lattice constant) 
tail with a $p_z$ character
on the halogen site in addition to a usual evanescent-wave like
part that exponentially decays away from the interface.
In the experiment\cite{Kiguchi}, 
while there is no information on Li sites 
since the element-selective NEXAFS was tuned for Cl in that 
experiment, 
an unoccupied state on Cl sites is observed near $E_F$.
They have a $p_z$ character, and spread with 
a decay length of a few \AA\. These properties are totally
consistent with the present results.

\begin{figure}
\begin{center}
\leavevmode\epsfysize=70mm \epsfbox{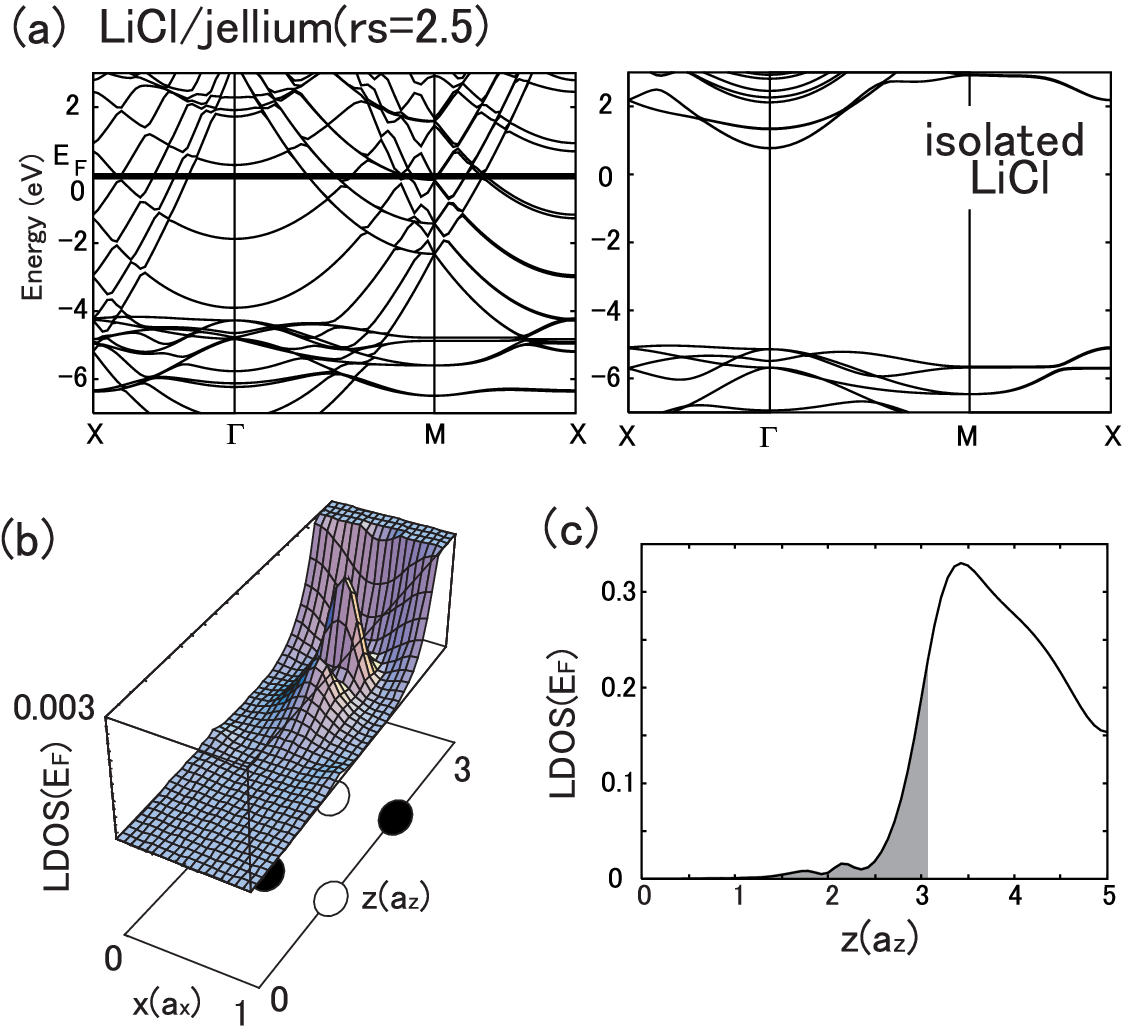}
\caption{A plot similar to Fig.\ref{rs25}, for a system where 
LiCl double layers sandwich the jellium having $r_s=2.5$.
}
\label{rs252}
\end{center}
\end{figure}

To confirm that the penetration depth of the tail of MIGS 
is as large as a few \AA,
we have also calculated a larger system, where 
LiCl double layers sandwich the jellium.  
From the band structure and the LDOS in Fig.\ref{rs252} 
we can see that the MIGS do not penetrate into the
second layer of LiCl.

\begin{figure}
\begin{center}
\leavevmode\epsfysize=70mm \epsfbox{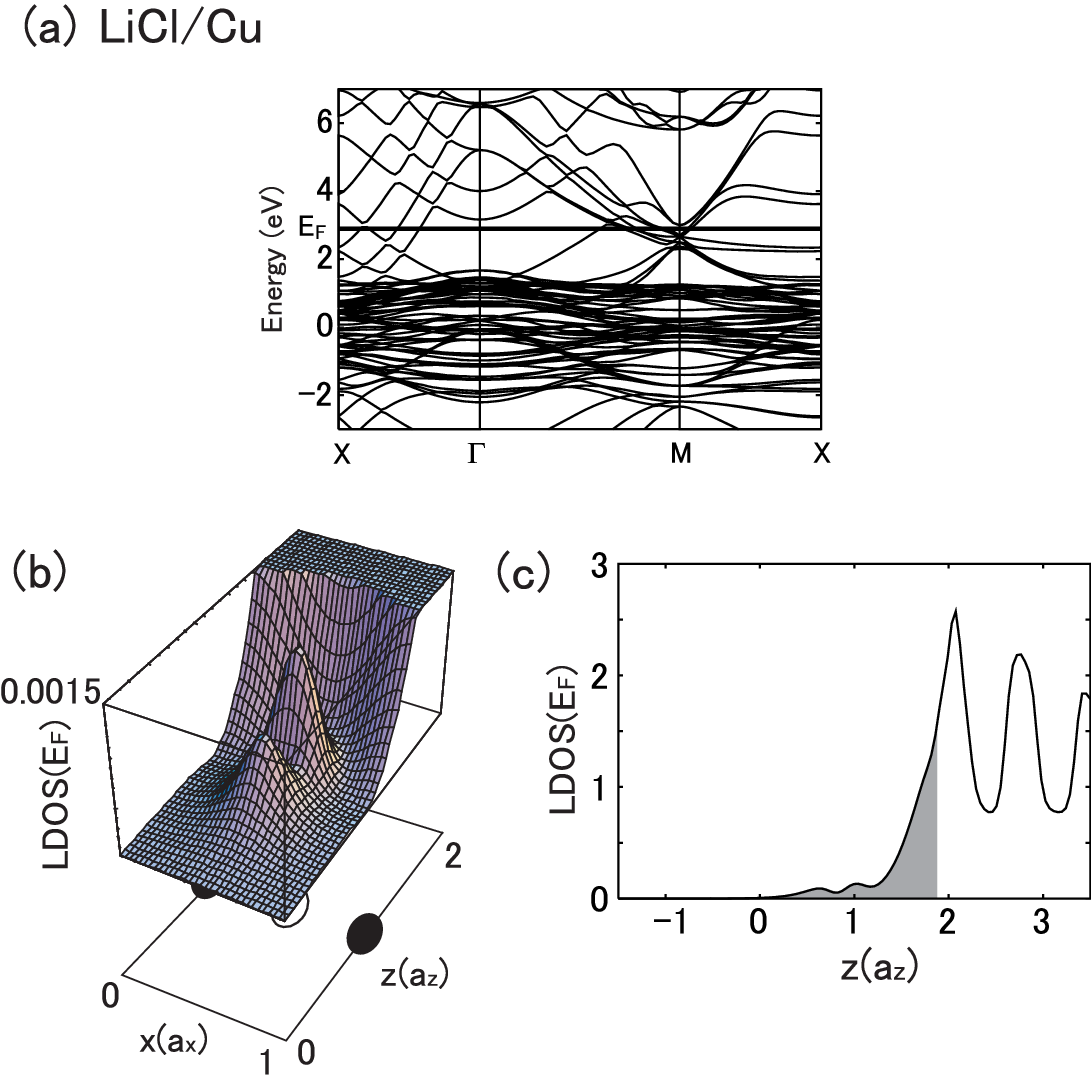}
\caption{A plot similar to Fig.\ref{rs25}, when the jellium is 
replaced with five layers of Cu.
}
\label{LiClCu}
\end{center}
\end{figure}

If we now replace the jellium 
with (five layers of) Cu in Fig.\ref{LiClCu}, 
we can see that the LDOS for the LiCl/Cu has a long tail 
just as in the result for the jellium model, so that
the characteristic feature of MIGS should not be
an artifact of the jellium model.

\begin{figure}
\begin{center}
\leavevmode\epsfysize=35mm \epsfbox{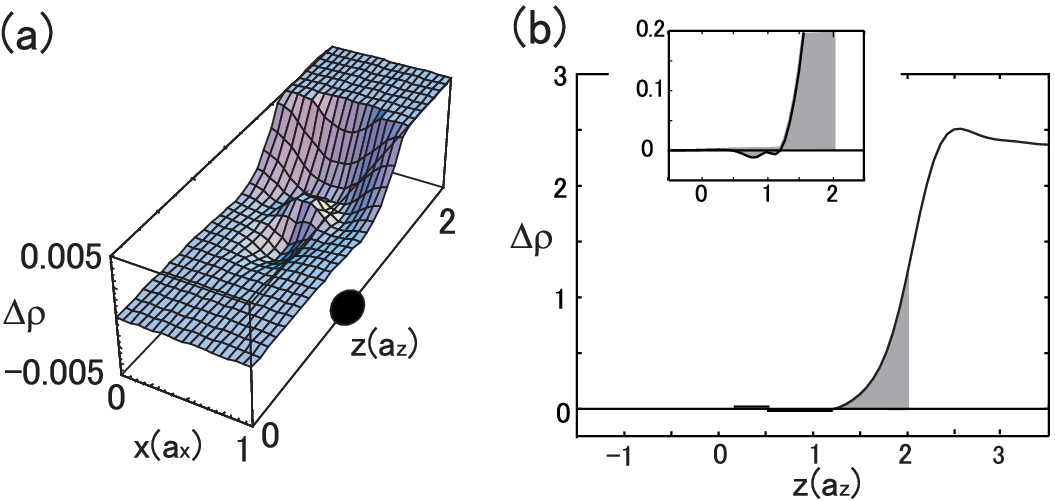}
\caption{(a) The difference in the valence charge density, 
$\delta \rho\equiv \rho_1-\rho_2$, integrated over $y$-direction, 
where $\rho_1$ is the density for LiCl+jellium and 
$\rho_2$ for the isolated LiCl.
(b) $\delta \rho$ integrated over both 
$x$ and $y$ directions. Inset is a blow-up 
for $-0.05<\delta\rho<0.05$. 
Shaded regions indicate the insulator region.
}
\label{charge}
\end{center}
\end{figure}

Why MIGS have amplitudes only on Li sites can be explained 
as follows. In general, MIGS can be divided into two groups: 
one having a conduction-band character and the other 
having a valence-band character in the insulator side of the interface. 
Noguera {\it et al}\cite{Noguera} have introduced the energy 
$E_{\rm ZCP}$, above which MIGS have a valence-band character.
While the charge neutrality in the insulator side is satisfied
when $E_{\rm ZCP}$ is equal to $E_F$, charge transfer from 
the insulator to the metal occurs if $E_{\rm ZCP} > E_F$.
In Fig.\ref{charge}, we plot the difference in the valence-charge density 
$\delta \rho$ between the LiCl+jellium and the 
isolated LiCl. We can see that, while $\delta \rho$ is positive 
for most of the insulator region ($0<z<2$), 
it is negative around the Cl site.
Thus $E_{\rm ZCP}$ should be higher than $E_F$ in this system,
and MIGS should have a tail on the anion sites.

\begin{figure}
\begin{center}
\leavevmode\epsfysize=70mm \epsfbox{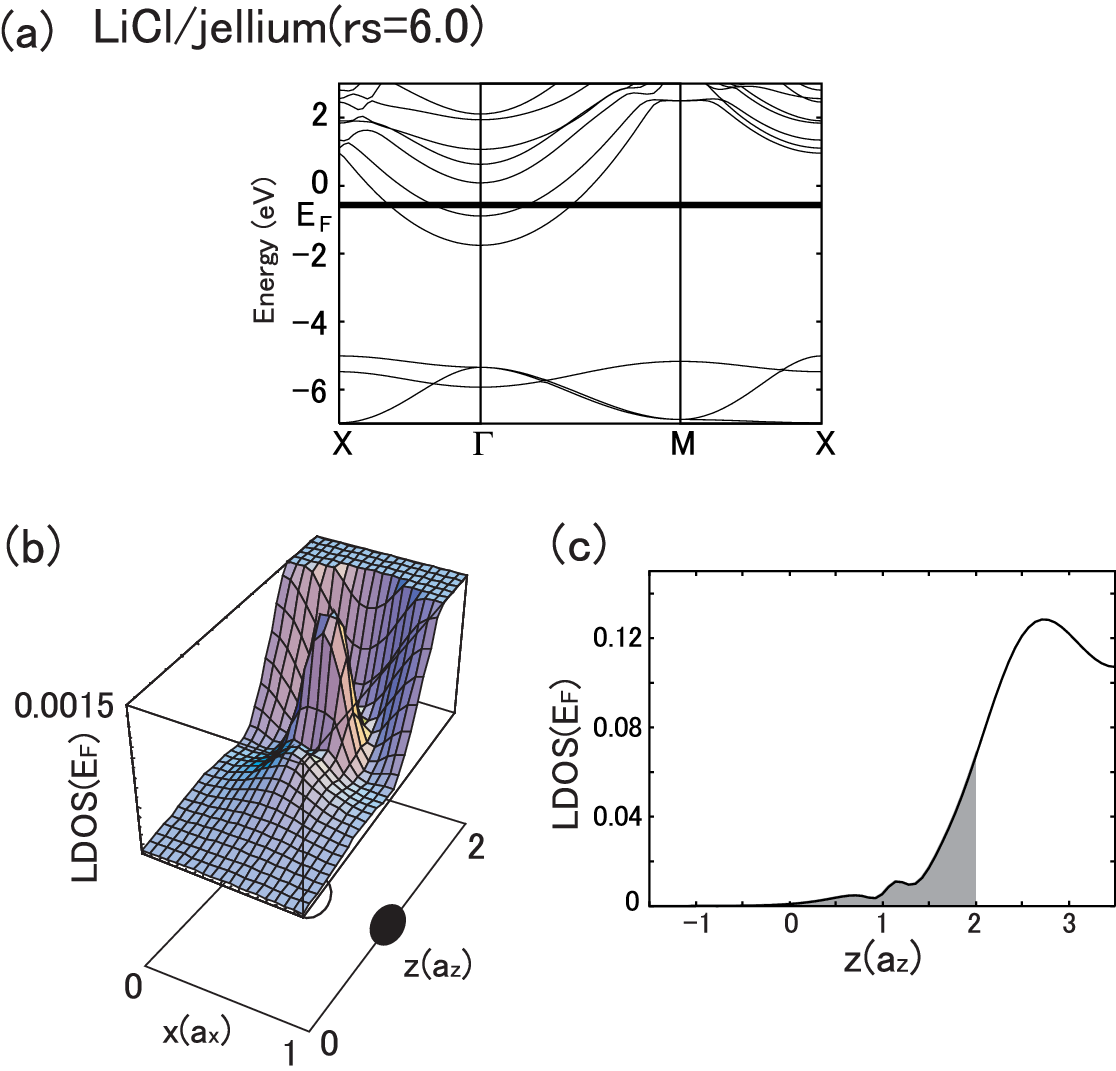}
\caption{A plot similar to Fig.\ref{rs25} when 
$r_s$ is increased from $2.5$ to $6.0$.
}
\label{rs6}
\end{center}
\end{figure}

Let us move on to the $r_s$ dependence.
In Fig.\ref{rs6}, we show the result when $r_s$ is 
increased from $2.5$ to $6.0$. 
We can see that, although LDOS at the interface is 
smaller than that for $r_s=2.5$, the decay length
for $r_s=6$ is similar to that for $r_s=2.5$.
Thus we may expect that penetration depth does not 
sensitively depend on the nature of the metal.

\subsection{Other alkali halides}
\begin{figure}
\begin{center}
\leavevmode\epsfysize=70mm \epsfbox{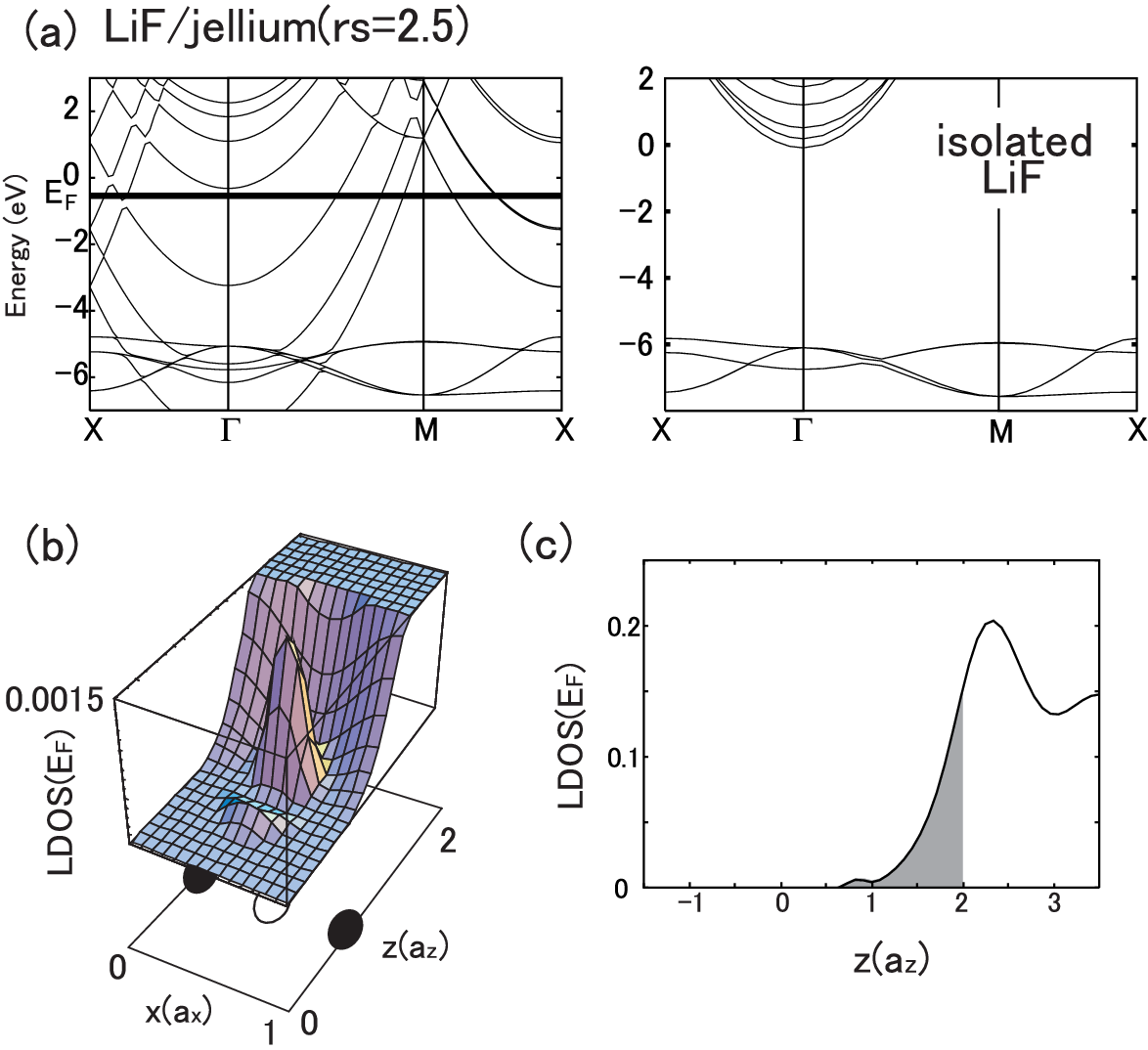}
\caption{A plot similar to Fig.\ref{rs25} with LiCl 
replaced by LiF. 
}
\label{LiF}
\end{center}
\end{figure}

\begin{figure}
\begin{center}
\leavevmode\epsfysize=70mm \epsfbox{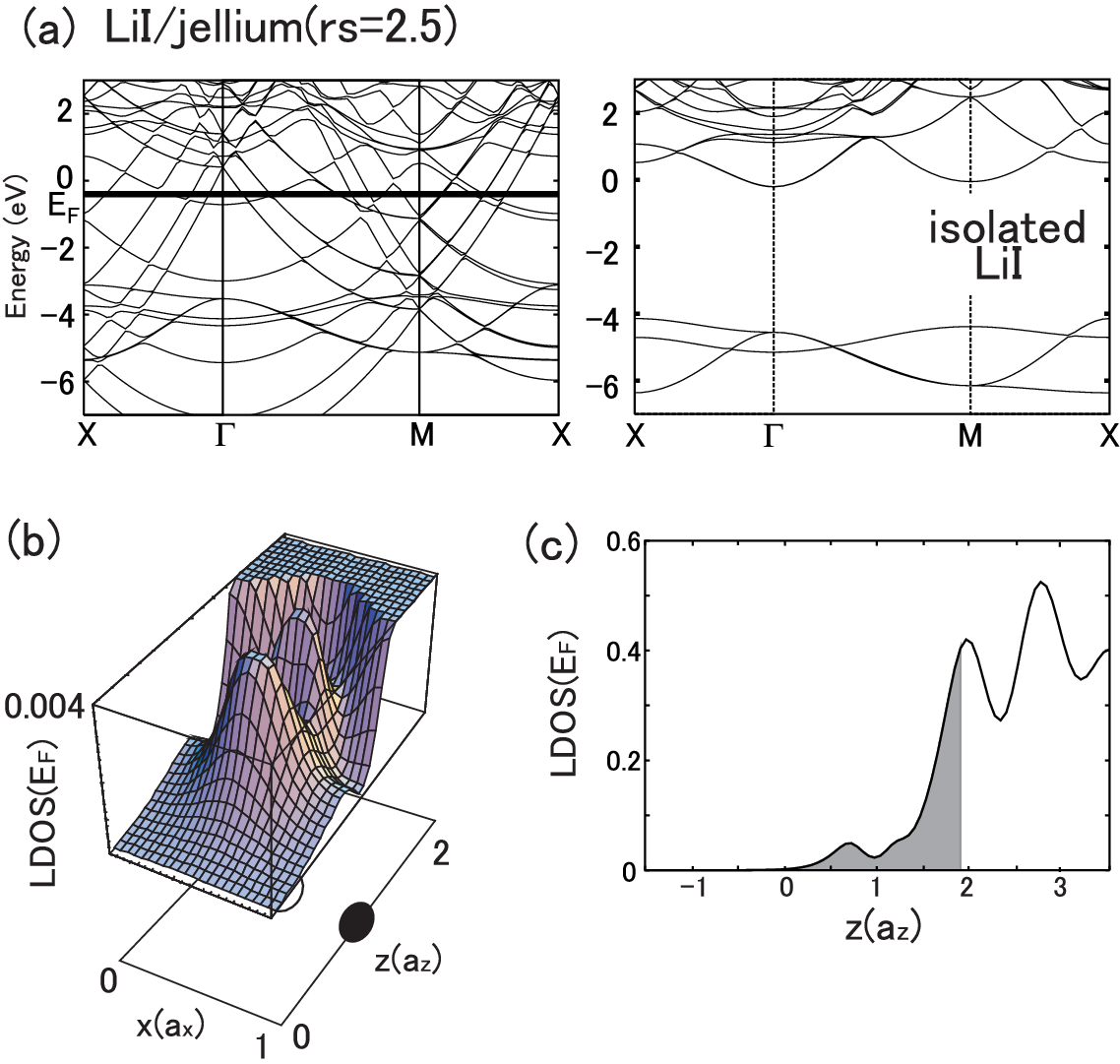}
\caption{A plot similar to Fig.\ref{rs25} with LiCl 
replaced by LiI.
}
\label{LiI}
\end{center}
\end{figure}

\begin{figure}
\begin{center}
\leavevmode\epsfysize=70mm \epsfbox{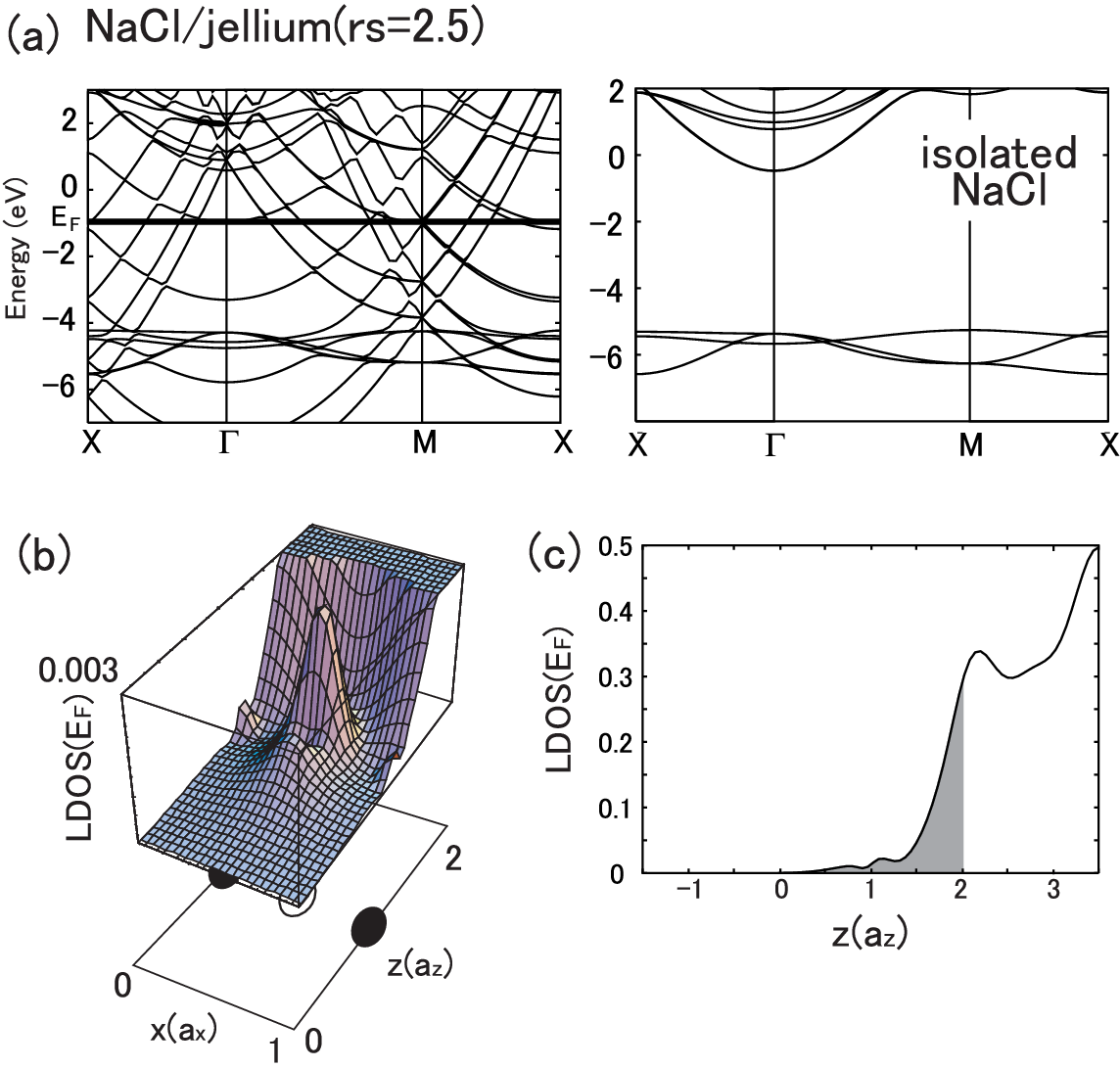}
\caption{A plot similar to Fig.\ref{rs25} with LiCl 
replaced by NaCl.
}
\label{NaCl}
\end{center}
\end{figure}

Let us move on to the cases of other alkali halides.
We have performed calculations for LiF, LiI and NaCl. 
The results are shown in Fig.\ref{LiF} for LiF, Fig.\ref{LiI} for LiI,
and Fig.\ref{NaCl} for NaCl.  The results are similar to 
that for LiCl(Fig.\ref{rs25}).  Namely, 
MIGS have a long tail on anion sites
with a penetration depth $\lambda$ as large as the 
layer spacing $a_z=a/2$.
The lattice constants are
$a=4.09$ \AA\ for LiF, 5.13 \AA\ (LiCl), 5.95 \AA\ (LiI) 
and 5.63 \AA\ (NaCl),
and $\lambda$ for LiI is indeed the largest (1.5 times as 
large as that for LiF).
Note that anions with higher electron affinity have 
smaller ionic radii in general, so that alkali halides with
largeer band gap have smaller lattice constants.

\section{Possibility of exciton mediated pairing}
One of the most fascinating possibilities which should 
be explored for heterointerfaces is, we believe, the exciton-mediated
superconductivity. The exciton mechanism
has been originally proposed by Little\cite{Little64} for
metallic, quasi-one dimensional spines of conducting electrons 
to which dielectric organic molecules are conceived 
to be attached chemically.
Subsequently, two-dimensional version of this mechanism 
has been proposed\cite{Ginzburg64,ABB,Ginzburgbook}, 
where a metallic film covered by a dielectric coating
was considered.

To realize the exciton mechanism in general, carriers 
have to coexist and strongly interact with the excitons.
Thus the penetration length must be sufficiently large,
which is why it has been believed that the 
band gap of the coating material should not be too large.
On the other hand, 
Inkson and Anderson\cite{Inkson} have pointed out that, 
if we adopt the
model dielectric function for typical semiconductors, 
exciton-electron coupling becomes small, 
partially because the exciton has a large energy dispersion
so that the phase space over which the mechanism works for 
the formation of paring is restricted.
In fact, in Ref.\cite{ABB} the energy dispersion 
of the exciton is ignored.  In other words, 
they have considered a Frenkel exciton, 
which is usually unrealistic for (narrow gap) semiconductors. 

Now, the present result for MIGS at metal/insulator
interfaces can reconcile the very dilemma. Namely, 
while a conventional wisdom tells that 
the penetration depth should be 
extremely small for metal/wide-gap insulator interfaces, 
the present result suggests that the depth is of the order 
of the lattice constant for the MIGS, so that a substantial portion of 
the carriers can interact with the exciton in the insulator side.
Thus it becomes an intriguing problem to estimate the electron-exciton
coupling for the alkali halide/metal system, so we do this 
following the argument of ref.\cite{ABB}.

According to ref.\cite{ABB}, the exciton-electron coupling 
$\lambda_{\rm ex}$ is estimated as 
$s \gamma^2 (D/L) \mu (\omega_p/\omega_g)^2$, where 
$s$ is the screening factor for the Coulomb interaction, 
$\gamma$ the ratio between the LDOS at the interface and that for the bulk, 
$D$ the average depth of penetration, 
$L$ the thickness of the metal film, 
$\mu$ the screened Coulomb interaction averaged over
the Fermi surface multiplied by the density of states at $E_F$, 
$\omega_p$ the electronic plasma frequency, 
and $\omega_g$ the band gap of the insulator.
If we assume that the dielectric function can be approximated as
\[
\varepsilon(\omega)=1 - 
\frac{(\varepsilon_{\rm optical} - 1)\omega_0^2}
{\omega^2-\omega_0^2},
\]
where $\omega_0 \sim \omega_g$ is the typical exciton energy, and 
$\varepsilon_{\rm optical}$ the optical dielectric constant, 
then $\lambda_{\rm ex}$ becomes $s \gamma^2 (D/L) \mu 
(\varepsilon_{\rm optical} - 1)$.

Let us first look into the factor $\gamma D/L$.
This factor is, in the simplified model due to Bardeen, 
the fraction of the time over which 
the metal electrons spend in the insulator.  
In the more realistic model 
we have at hand, we can more precisely 
integrate the LDOS in the insulator region ($\equiv \rho_i$, 
shaded region in Fig.\ref{rs25}(c) etc) 
since MIGS do not possess a monotonic (exponential) LDOS, 
and we can then divide the quantity by that for the metal 
region ($\equiv\rho_m$) to obtain the factor corresponding 
to $\gamma D/L$.  
Namely, the factor should read
$b\equiv 2\rho_i/\rho_m$,
in place of $\gamma D/L$, where $a_z=a/2$ (see Fig.\ref{unit}).
The present model is a sandwich structure where 
the metallic layer is covered by the dielectric coatings 
from both sides, so that the factor 2 appears in r.h.s. 
Here we set that the thickness of metallic layer to be 10 \AA. 
The resulting values are $b =0.08$ (for LiF), 0.18(LiCl), 
0.16(LiI), and 0.12(NaCl).

On the other hand, $(\varepsilon_{\rm optical} - 1)$ are
0.92, 1.68, 2.4 and 1.33 for LiF, LiCl, LiI and NaCl,
respectively\cite{Lines90}.
Therefore, if we adopt the same values for $s(=1/2)$, 
$\mu(=1/3)$, $\gamma(=1/2)$ as in Ref.\cite{ABB},
$\lambda_{\rm ex}$ can become as large as 0.06 for LiI, 
while we have $\lambda_{\rm ex}=0.01$(LiF), 
0.05( LiCl), and 0.03(NaCl). 

If we adopt weak-coupling conventional superconductors, 
e.g. Al, as the metal, the superconducting
transition temperature can be roughly estimated\cite{Ginzburgbook,Tc} as 
\[
T_c = \theta \exp\left(-\frac{1}{(\lambda_{\rm ph}+\lambda_{\rm ex}
-\mu^*)}\right),
\]
where $\lambda_{\rm ph}$ is the electron-phonon coupling, 
$\lambda_{\rm ex}$ the electron-exciton coupling, and 
$\mu^*$ the renormalized $\mu$.   
So $\theta$ is roughly expressed as
\[
\log \theta \propto \int (\log \omega) \left[\rho_{\rm ex} (\omega)
+\rho_{\rm ph} (\omega)\right]\frac{d\omega}{\omega},
\]
where $\rho_{\rm ph}$ ($\rho_{\rm ex}$) is the density of 
phonon (exciton) states. In general, $\rho_{\rm ph}$ 
has a peak in a low energy region, so that, when phonons and 
excitons coexist, the contribution of
the phonon part to $\theta$ should be dominant than that of the 
exciton part. 
Thus $\theta$ basically scales with that of phonon, 
but the exciton contribution on top of that can enhance 
the $T_c$ significantly.  
For example, when $\lambda_{\rm ph}\simeq 0.4$, $\mu^*\simeq 0.05$, and 
$\lambda_{\rm ex}\simeq 0.1$, 
$T_c$ becomes 1.6 times greater than that in the bulk. 
Thus we conclude that heterointerface between metal 
and LiI or LiCl can be a promising ground for detecting the 
existence of exciton-enhanced superconductivity.

In addition, the proximity of different systems enables 
us to expect entirely different mechanism. 
Namely, two of the present authors\cite{Kuroki} have
earlier proposed that a class of repulsively interacting systems 
that consist of a carrier band and an insulating band
can become superconducting, where the system can effectively 
mapped to an {\it attractive} Hubbard model.  The point is, 
while the conventional boson-exchange pairing arises due to 
a retarded attraction, 
superconductivity in the carrier-insulator model occurs due to a 
{\it nonretarded} attraction. 
In their model, the metallic band is assumed to interact only with 
the anion sites (or to interact only with the cation sites).  
At the interface between
an alkali halide and a metal, MIGS have their amplitudes 
on anion sites, so that the MIGS should interact primarily with 
anion sites, which just corresponds to the situation considered 
in Ref.\cite{Kuroki}.
The detailed study on the combined effect of the 
nonretarded attraction (which decreases $\mu^*$)
and the exciton effect (which increases $\lambda$j
is an interesting future problem.

\section{Summary}
We have studied the
electronic properties of metal-induced gap states(MIGS) at interfaces 
between various alkali halides and a metal.
We have performed first-principles density functional calculation
to show that (i) in addition to a usual evanescent-wave component, 
MIGS have a tail on halogen sites with 
$p_z$-like character, whose penetration depth 
is as large as the lattice constant of the alkali halide,
(ii) while $\lambda$ does not significantly depend on the carrier 
density of jellium, it depends on the energy gap of alkali halides,
with $\lambda_{\rm LiF} < \lambda_{\rm LiCl} < \lambda_{\rm LiI}$.
We have estimated the electron-exciton coupling, and have found that
we may envisage that insulator/metal interfaces such as those 
discussed in the present study may provide a possible ground for 
exciton-mediated superconductivity. 

\section{Acknowledgment}
We would like to thank K. Saiki, M. Kiguchi and
D. R. Jennison for fruitful discussions. 
The GGA calculation was performed with 
TAPP (Tokyo Ab-inito Program Package), where RA and YT would like 
to thank Y. Suwa for technical advices.
Numerical calculations 
were performed on SR8000 in ISSP, University of Tokyo.
This research was partially supported by the Ministry of Education, 
Science, Sports and Culture, Grant-in-Aid for Creative Scientific
Research.

\end{multicols}
\end{document}